# Effects of the applied fields' strength on the plasma behavior and processes in E×B plasma discharges of various propellants: II. Magnetic field

M. Reza*[1], F. Faraji*, A. Knoll*

*Plasma Propulsion Laboratory, Department of Aeronautics, Imperial College London, London, United Kingdom

**Abstract**: Following the discussions in part I of the article, we present in this part II the effects of the magnetic field intensity on the properties of the plasma discharge and the underlying phenomena for different propellant's ion mass. The plasma setup represents a perpendicular configuration of the electric and magnetic fields, with the electric field along the axial direction and the magnetic field along the radial direction. The magnetic field intensity has been changed from 5 to 30 mT, with 5 mT increments. The propellant gases are xenon, krypton, and argon. The simulations are carried out using a particle-in-cell (PIC) code based on the computationally efficient reduced-order PIC scheme. Similar to the observations in part I, we show that, across all propellants, the variation in the intensity of the magnetic field yields two distinct regimes of the plasma, where either the Modified Two Stream Instability (MTSI) or the Electron Cyclotron Drift Instability (ECDI) are present. Nonetheless, a third plasma regime is also observed for cases with moderate values of the magnetic field intensity (15 and 20 mT), in which the ECDI and the MTSI co-exist with comparable amplitudes. This described change in the plasma regime becomes clearly reflected in the radial distribution of the axial electron current density and the electron temperature anisotropy. Contrary to the effect of the electric field magnitude in part I, we observed here that the MTSI is absent at the relatively *low* magnetic field intensities (5 and 10 mT). At the relatively high magnitudes of the magnetic field (25 and 30 mT), the MTSI becomes strongly present, a long-wavelength wave mode develops, and the ECDI does not excite. An exception to this latter observation was noticed for xenon, for which the ECDI's presence persists up to the magnetic field peak value of 25 mT.

## Section 1: Introduction

E×B plasma discharges, which are of high applied relevance across the industries, from the manufacturing to the aerospace, present intricate physics that has gathered the attention of the scientific community for decades. The plasma in these configurations is characterized by a vast array of underlying multi-scale, multi-dimensional, and multi-physics processes [1][2]. The resulting highly complex nature of the plasma has impeded achieving a comprehensive knowledge basis of the phenomena in the E×B plasmas despite years of effort and a large body of research [1]-[6]. An enabling effort to achieve this full understanding consists of extensive parametric studies over wide ranges of plasma conditions and in controlled simulation environments to isolate the effects of various physical factors and mechanisms on the underlying microscopic processes as well as the global macroscopic behavior of the plasma. The experimental works, informed by these simulations, are also of high importance in complementarity to such an effort since they enable the validation of the numerical observations and provide further insights from real-world devices and settings. The existing rich literature can indeed guide the joint numerical-experimental endeavors by highlighting the phenomena and/or the range of plasma parameters/configurations that require further in-depth analyses and more focused investigations.

The available high-fidelity plasma simulation codes, such as the ones based on the kinetic particle-in-cell (PIC) method [7], have so far proven to be highly useful to provide insights into the complicated physics of the E×B discharges. However, their enormous computational cost does not allow for extensive parametric investigations that are required to study in detail the characteristics of the multi-scale, multi-dimensional phenomena over broad parameter spaces as well as the involved dynamics over rather extended time frames. The reduced-order PIC scheme [8]-[12], developed and extensively verified at Imperial Plasma Propulsion Laboratory (IPPL), serves as a game-changer in this respect since it enables simulations of high fidelity, that closely resemble the results of traditional PIC codes, to be performed at computational costs reduced by several factors to an order of magnitude [8][9][12].

Equipped with this cost-efficient numerical tool and following on the investigations presented in part I [13] of this article, we focus in this part II on the effects that the intensity of the applied magnetic field in our adopted radial-azimuthal plasma configuration have on the development, characteristics, and the impacts of the ECDI and the MTSI modes in plasma discharges of xenon, krypton, and argon.

---

[1] **Corresponding Author** (m.reza20@imperial.ac.uk)

It is important to note that, in the context of E×B plasma technologies, the magnetic field intensity represents a design parameter. Particularly in the case of Hall thrusters, which we have chosen as a proving ground for our analyses here, the topology of the applied magnetic field and its magnitudes within the domain remain invariant over time and through the operation of the device. However, the magnetic field typically features notable spatial variations by design, with conventionally large gradients existing in the axial direction along the thrusters' discharge channel centerline [14][15].

To the best of our knowledge, the present effort that aims to systematically assess in a radial-azimuthal setup the distinct influences of the magnetic field intensity on the instability spectra and the consequent variation in the properties and the behavior of plasma discharges of multiple propellants is unique and un-attempted within the existing literature. In fact, prior to this work, the effects of the magnetic field intensity in a similar context to our paper have been numerically only assessed in an axial-azimuthal Hall-thruster-representative configuration and with xenon propellant [16][17]. The research reported in Refs. [16][17] investigated the spatial variations in the characteristics of the ECDI and its transition dynamics to the Ion Acoustic Instability [16][18] in the presence of various magnetic field magnitudes.

From an experimental perspective, even though the authors are unaware of any effort with a similar scope to that of this article, there have been multiple publications that examined for E×B plasma discharges the influence of the magnetic field intensity on the large-scale plasma instabilities such as the spoke mode [19][20], the electron species' properties [21], and their cross-field transport characteristics [22][23].

## Section 2: Description of the simulations' setup

The simulations' domain is a 2D Cartesian plane representing the radial-azimuthal coordinates of a Hall thruster. In the adopted coordinate system, $x$, $y$, and $z$ indicate the radial, axial, and azimuthal directions, respectively. The details of the simulations' setup follow identically the description provided in part I of the article [13]. The only exception is that, for the simulations in this paper, the strength of the applied radial magnetic field ($B_x$) is varied across a range of values, which are summarized in Table 1. The axial electric field remains constant at $10\ kVm^{-1}$ in all cases. This is the same value of the electric field as that adopted in the radial-azimuthal benchmark case in Ref. [24].

| Case No. | $B_x\ [mT]$ | $B_x/B_0$ | Propellant |
|---|---|---|---|
| 1 | 5 | 0.25 | Xe, Kr, Ar |
| 2 | 10 | 0.5 | Xe, Kr, Ar |
| 3 | 15 | 0.75 | Xe, Kr, Ar |
| 4 | 20 | 1 | Xe, Kr, Ar |
| 5 | 25 | 1.25 | Xe, Kr, Ar |
| 6 | 30 | 1.5 | Xe, Kr |

Table 1: List of the studied simulation cases, summarizing the value of the radial magnetic field ($B_x$) and the ion species used in each case. $B_0$ refers to the magnetic field intensity in the baseline benchmark setup [24].

## Section 3: Results and discussion

In this section, we present and discuss the results of the analyses performed to investigate the impact of the intensity of the radial magnetic field on the dynamics of the plasma discharge with various propellants (gases). Our focus is on studying the variation with $B_x$ of the dominant instability modes and their effects on the plasma. Following the same order of discussions as that in part 1 [13], we begin by presenting the variation in the plasma properties' profiles. We then discuss the instabilities' characteristics as deduced from the Fast Fourier Transform (FFT) analyses of the azimuthal fluctuations and complemented by a more comprehensive Dynamic Mode Decomposition (DMD) analysis. Moreover, for the xenon propellant as an example, we discuss the various contributions to the electrons' cross-field transport including the roles of the instabilities, the viscosity, and the convective inertia force terms. This is followed by analyzing the instabilities' impacts on the particles' velocity distribution functions across the radial and azimuthal components of the velocity for various propellants.

## 3.1. Variation in the plasma properties distribution

Figure 1(a) shows the time-averaged radial profiles of the ion number density for different simulation cases, and Figure 1(b) presents the variation of the peak value of the ion number density profiles with $B_x$. We observe a consistent trend of decreasing density with increasing magnetic field intensity, with Xe and Ar consistently exhibiting the highest and the lowest densities, respectively. Furthermore, beyond 15 mT, the variation in the density becomes very small, and the average density evolution with $B_x$ appears to reach an asymptotic value.

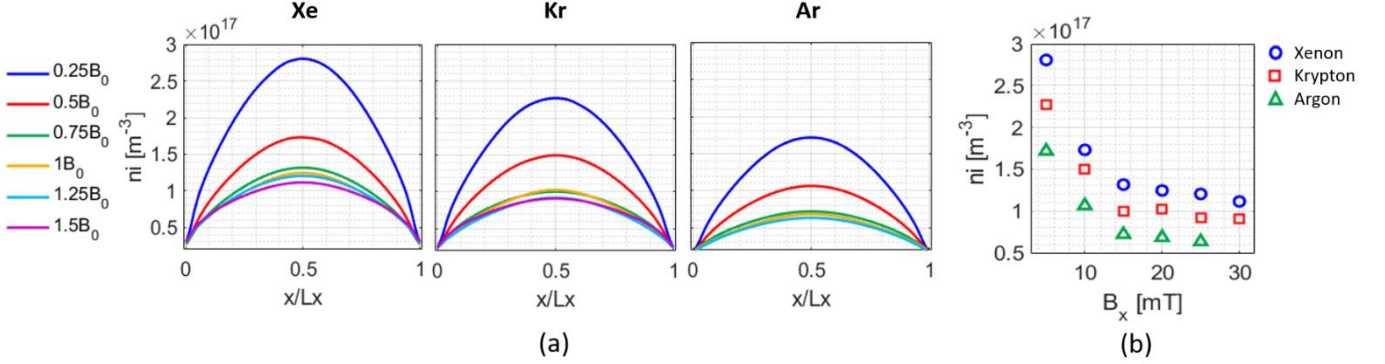

Figure 1: (a) Radial profiles of the ion number density ($n_i$) averaged over 20-30 $\mu s$ from the radial-azimuthal simulations with various radial magnetic field intensities and propellants. (b) Variation vs $B_x$ of the ion number density's peak value along the radial direction for the three studied propellants.

The radial distributions of the electron temperature ($T_e$), and the radial and the azimuthal electron temperatures ($T_{ex}$ and $T_{ez}$) as well as the average value of these properties over the radial extent of the domain are plotted in Figure 2.

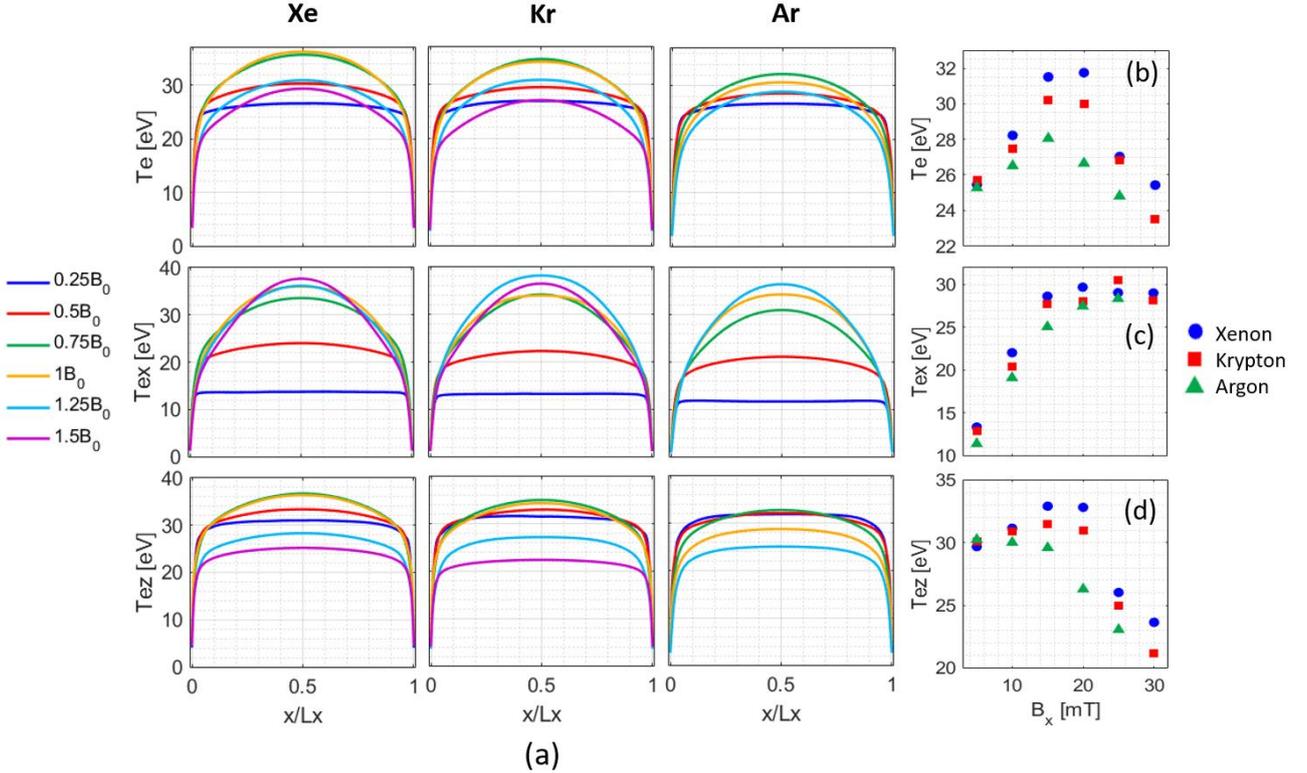

Figure 2: (a) Radial profiles of electron temperature ($T_e$), radial electron temperature ($T_{ex}$), and azimuthal electron temperature ($T_{ez}$), averaged over 20-30 $\mu s$ from the radial-azimuthal simulations with various radial magnetic field intensities and propellants. (b) – (d) Variation vs $B_x$ of the radially averaged $T_e$ (plot (b)), $T_{ex}$ (plot (c)), and $T_{ez}$ (plot (d)) for the three studied propellants from the quasi-2D simulations.

From these plots, we notice that $T_e$ exhibits an increasing trend within the $B_x$ range of 5 to 15 mT for all propellants. For Xe and Kr, the $T_e$ remains nearly constant between 15 and 20 mT but decreases afterwards toward higher $B_x$ values. However, for Ar, $T_e$ consistently drops beyond 15 mT. The $T_{ez}$ follows a similar variation with $B_x$ as that of the $T_e$ in the cases of Xe and Kr propellant. Nonetheless, for Ar, the azimuthal electron temperature does not vary significantly for the magnetic fields between 5 to 15 mT.

The $T_{ex}$ for all propellants increases with the magnetic field up to $B_x = 15$ mT after which it remains almost constant. We can readily notice by comparing the $T_{ex}$ vs $B_x$ plot against the $n_i$ vs $B_x$ plot of Figure 1(b) the inverse correlation between the radial electron temperature, which determines the particles' thermal flux to the walls, and the plasma density. This inverse correlation arises due to the temporal constancy of the imposed injection source in the simulations as elaborated on in part I [13]. Nevertheless, it is emphasized that the extent to which the observed correlation may hold in more realistic scenarios needs to be investigated in a setup with self-consistent ionization process for which the axial fluxes of the particles are also self-consistently resolved.

In part I [13], we exploited the knowledge of the different influences that the MTSI and the ECDI have on the electron heating and the electron cross-field transport to distinguish the dominant instability among the ECDI and the MTSI across various simulation cases prior to analyzing the FFT/DMD spectra of the waves.

We introduced two parameters, namely, the average of $T_{ex}/T_{ez}$ ratio over the radial extent, and the radial profile of the normalized axial electron current density ($J_{ey}/J_{ey,mean}$), as two indicators for the above purpose. In this regard, as the ECDI causes the electron heating primarily in the azimuthal direction [25][26], whereas the MTSI heats up the electrons mostly in the radial direction [27][28], the ratio of the radial-to-azimuthal electron temperature can be an indirect indicator of the relative significance of the MTSI and the ECDI modes in a simulation. In addition, as observed in part I [13] and in Ref. [29], the electron axial current induced by the MTSI is mainly concentrated near the walls as opposed to that carried by the ECDI, which is mostly focused in the central region of the domain.

Accordingly, looking at Figure 3, we can see that, for high values of the magnetic field (25 and 30 mT), $\frac{T_{ex}}{T_{ez}} > 1$ and $J_{ey}$ is mainly the largest near the walls for all propellants. In case of Xe with $B_x = 25$ mT, the spikes in the $J_{ey}$ near the walls are more pronounced than for other propellants. We can, therefore, deduce that, in these cases, the MTSI is stronger than the ECDI. In the next Section, we demonstrate that the ECDI is, in fact, absent at $B_x = 30$ mT for Xe and Kr (Ar simulation not carried out) and at $B_x = 25$ mT for Kr and Ar.

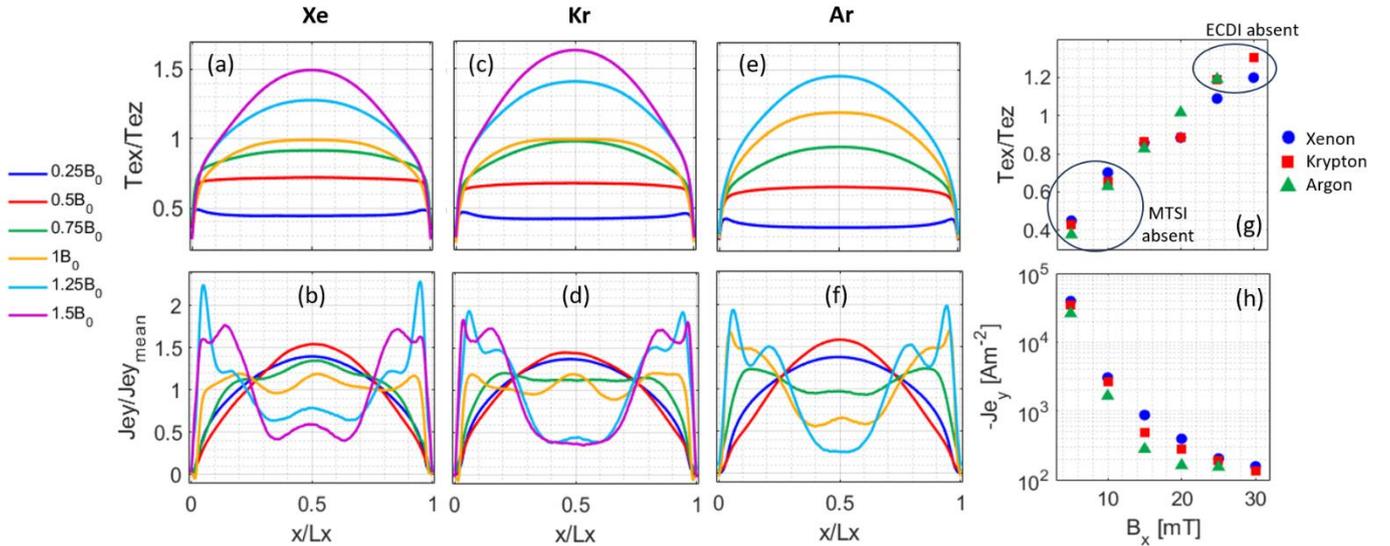

Figure 3: (a)-(f) Radial profiles of the radial-to-azimuthal electron temperature ratio ($T_{ex}/T_{ez}$) (top row) and the normalized axial electron current density ($J_{ey}/J_{ey,mean}$) (bottom row) averaged over 20-30 $\mu s$ from the quasi-2D simulations with various axial magnetic fields and propellants. (g) Variation vs $B_x$ of the radially averaged values of $T_{ex}/T_{ez}$ for the quasi-2D simulations with various propellants. (h) Variation vs $B_x$ of the radially averaged values of $-J_{ey}$ for the quasi-2D simulations with various propellants.

On the contrary, for low values of the magnetic field (5 and 10 mT), $\frac{T_{ex}}{T_{ez}} < 1$, and the $J_{ey}$ is maximum at the centerline for all propellants. This suggests that the ECDI must be the dominant mode in these conditions. We will see later that the MTSI does not indeed appear in these cases with low magnetic fields.

At the intermediate magnetic field values (15 and 20 mT), particularly for Xe and Kr, $\frac{T_{ex}}{T_{ez}} \approx 1$, and $J_{ey}$ has a more uniform distribution across the radial extent, which implies that both the MTSI and the ECDI develop with comparable intensities. Nevertheless, in the Ar simulation case with $B_x = 20$ mT, the MTSI seems to be slightly stronger than the ECDI as the ratio $\frac{T_{ex}}{T_{ez}}$ is moderately above unity and the $J_{ey}$ is larger toward the walls.

As another interesting observation, we can see from the radial profiles of the ion Mach number shown in Figure 4, that the ion sonic point for all propellants recedes from near the walls toward the plasma bulk as the magnetic field increases. This is an important point, which requires careful consideration for models which assume the Bohm's criterion to determine the edge of the sheath. Indeed, as pointed out in Ref. [30], in cases where the ion sonic point occurs inside the plasma bulk, the quasineutrality condition, for instance $\frac{|n_i - n_e|}{n_i} \approx 0.05$, can be a better criterion to delimit the edge of the sheath rather than Bohm criterion.

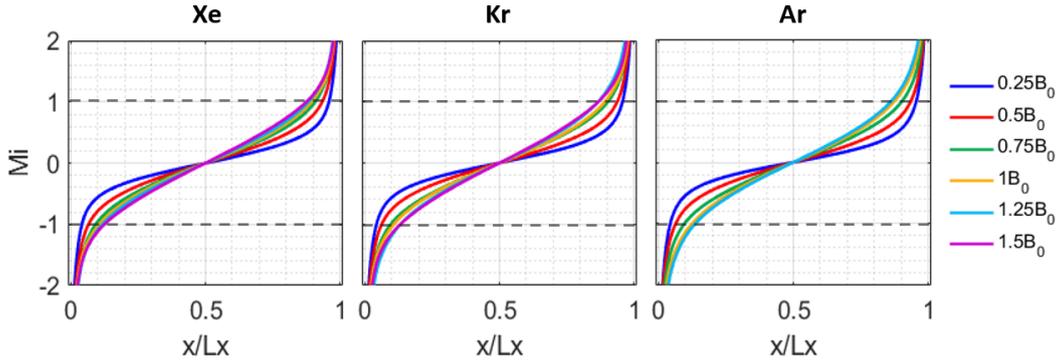

Figure 4: Time-averaged radial profiles of the ion Mach number over the time interval of $20 - 30 \ \mu s$ from the quasi-2D simulations with various magnetic field intensities and different propellants.

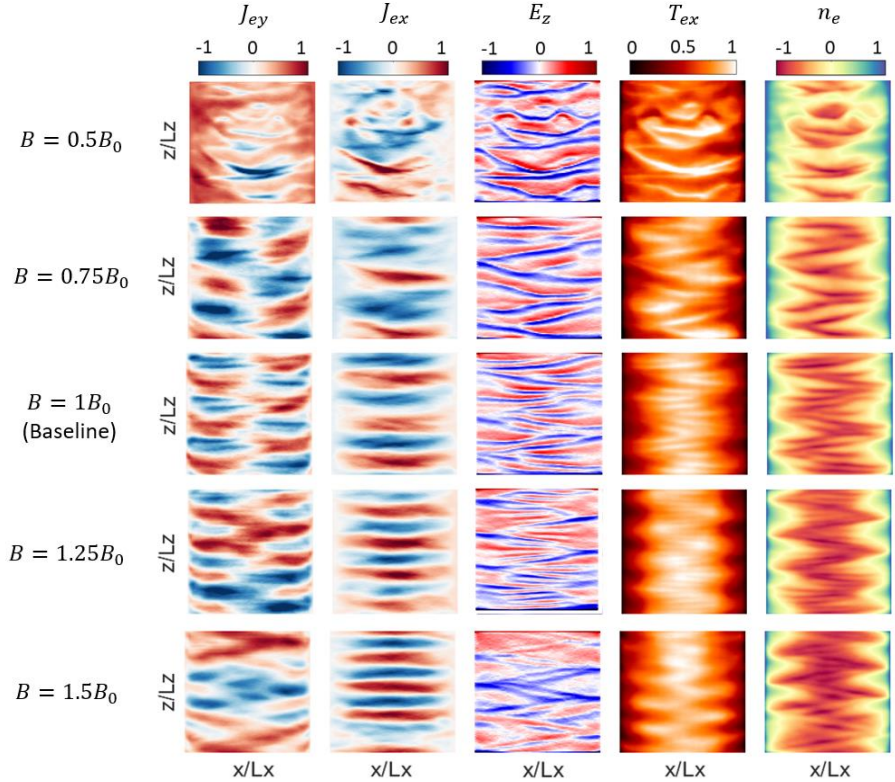

Figure 5: Comparison of the 2D snapshots of the normalized plasma properties at a time of local maximum of the radial electron temperature for various radial $B$-field intensities and the xenon propellant. The columns, from left to right, represent the axial and radial electron current densities ($J_{ey}$ and $J_{ex}$), the azimuthal electric field ($E_z$), the radial electron temperature ($T_{ex}$), and electron number density ($n_e$).

A sample set of snapshots of various plasma properties for Xe with different magnetic field intensities are presented in Figure 5. The snapshots in each $B_x$ case correspond to an instance of time when $T_{ex}$ is at its local maximum. The snapshots of the $0.25B_0$ case are not shown here, because they did not show distinguishable coherent structures as in the other cases. In the $0.5B_0$ case, the MTSI-like patterns do not appear, and only azimuthal waves are visible. Cases with $0.75B_0$ and $1B_0$ feature both azimuthal ECDI waves with short azimuthal wavelengths and the radial-azimuthal MTSI structures of longer azimuthal wavelength. The ECDI waves become less distinct in the $1.25B_0$ case and completely disappear in the $1.5B_0$ case. Furthermore, it is noteworthy that, at high magnetic field intensities ($1.25B_0$ and $1.5B_0$), a wave structure with a long azimuthal wavelength is seen to be developing as well.

In Figure 6, we compare the snapshots of the $J_{ey}$ and the $E_z$ for the three propellants and for various magnetic field intensities. From these snapshots, we can qualitatively observe the similarity of the 2D spatial distributions of these two plasma properties between different propellants at most magnetic field intensities. A disparity is noticed, however, for the $1.25B_0$ case, where the ECDI azimuthal waves are less pronounced for Kr than for Xe, whereas, for Ar, the ECDI waves seem to be absent completely. This points out to the fact that, for a comprehensive evaluation of the instabilities' characteristics, we need to assess the FFT and the DMD spectra of instabilities which will be presented in the subsequent section.

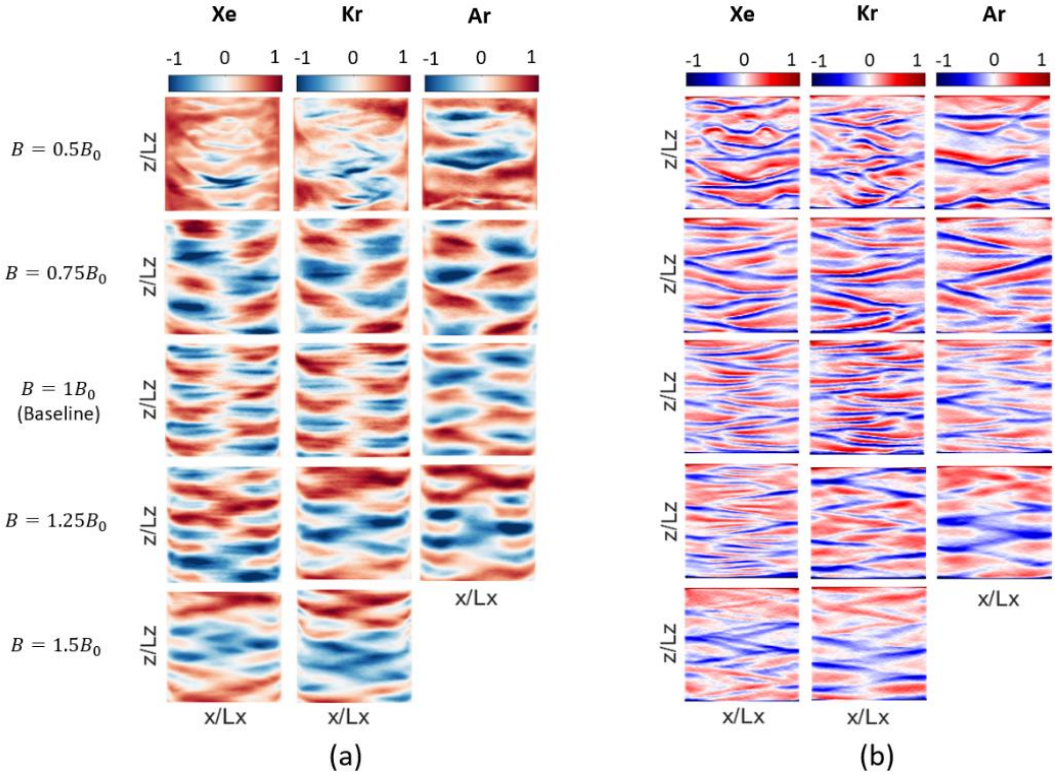

Figure 6: Comparison of the 2D snapshots of (a) normalized radial electron current density ($J_{ey}$), and (b) the normalized azimuthal electric field ($E_z$) at a time of local maximum of the radial electron temperature for various radial $B$-field intensities and the three propellants.

### 3.2. Variation in the characteristics and structure of azimuthal instabilities

The average spatial FFT spectra of the azimuthal electric field ($E_z$) are presented in Figure 7. The wavenumbers are nondimensionalized with respect to the azimuthal wavenumber of the ECDI's first harmonic, $k_0 = \frac{\Omega_{ce}}{V_{d_e}}$, in which $\Omega_{ce}$ and $V_{d_e}$ are the electron cyclotron frequency and the electron azimuthal drift velocity, respectively.

We can see from the plots in Figure 7 that, at the low magnetic field values ($0.25B_0$ and $0.5B_0$ cases) and for all propellants, only the ECDI mode exists with $\frac{k_z}{k_0} \approx 1$. In the $0.75B_0$ case, the FFT spectrum starts to show a peak near the MTSI's azimuthal wavenumber at $k_z/k_0 \approx 0.2$. In this case, the ECDI peak is comparatively lower in Kr than in Xe, and it reaches its lowest in Ar. In the case with $1B_0$, all propellants exhibit the presence of both the ECDI and the MTSI. Nevertheless, in the case of Ar, the intensity of the ECDI is nearly equivalent to that of the MTSI in contrast to the cases with Xe and Kr in which the ECDI is stronger than the MTSI.

In the high magnetic field range ($1.25B_0$ and $1.5B_0$ cases), the MTSI is clearly present with high amplitude together with another mode at a lower wavenumber and a comparable strength, which we refer to as the long-$\lambda$ mode. This long-$\lambda$ instability mode had been observed in our previous studies [29][31] with a similar wavelength and a frequency of approximately 1 MHz, particularly in the presence of strong radial magnetic field gradients, or significant Secondary Electron Emission (SEE), or at high number density conditions. The exact origin of this instability is still not fully understood; however, it is conjectured to be the result of an inverse energy cascade of shorter wavelength instabilities within the system, at least in certain conditions [29].

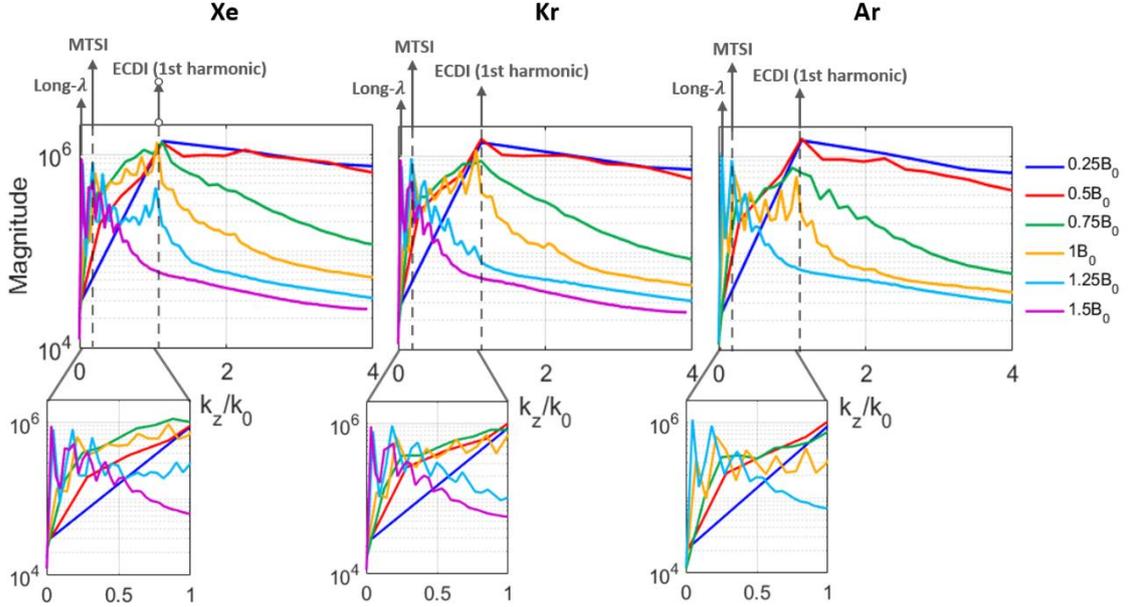

Figure 7: 1D spatial FFT plots of the azimuthal electric field signal from simulations with various values of $B_x$ and different propellants. The FFTs are averaged over all radial positions and over the time interval of 20-30 $\mu s$.

To determine the frequency spectrum of the instabilities, we look at the spatially averaged temporal FFT spectra of the $E_z$ signal shown in Figure 8 for various simulations. It should be noted from the FFT plots in Figure 8, however, that, in some cases, there are more than one peak within the expected frequency range of the instability of interest. Thus, as explained in part I [13], we have additionally used the DMD analysis [32] to distinguish the individual instability modes present in the system and to derive unambiguously their spatial structure together with their frequencies.

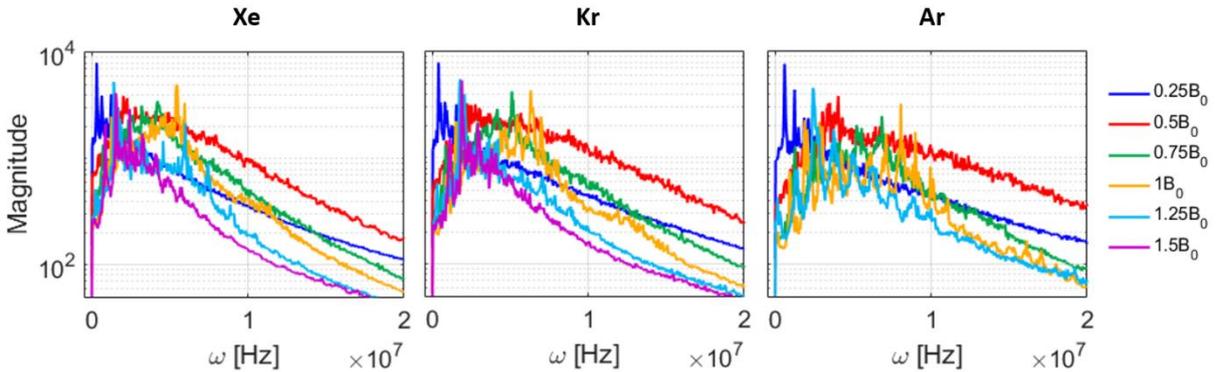

Figure 8: 1D temporal FFT plots of the azimuthal electric field signal from quasi-2D simulations with various magnitudes of the radial magnetic field and different propellants. The FFTs are averaged over all radial positions.

The results of the DMD analysis using OPT-DMD [33] for Xe are presented in Figure 9. In line with our conclusions from the spatial FFT analysis (Figure 7), the MTSI does not excite in the low-magnetic-field-conditions ($0.25B_0$ and $0.5B_0$ cases) because, in these two $B_x$ cases, there is no mode with radial and azimuthal structures resembling the MTSI. All DMD-discovered modes in these cases amount to azimuthal fluctuations which appear to be somewhat turbulent, and they are not as clean as the ECDI modes observed in the cases with larger magnetic field values. Another interesting characteristic of the ECDI and other azimuthal modes in these two cases is that the amplitude of the fluctuations for each mode appears to be nearly uniform across the domain (this is also the case for $0.75B_0$ case). On the contrary, in the higher $B_x$ conditions ($1B_0$ and $1.25B_0$), the intensity of the ECDI modes is higher near the centerline.

At the highest simulated magnetic field (the $1.5B_0$ case), the DMD does not discover any mode with purely azimuthal oscillations representative of the ECDI, which is again consistent with the absence of a peak near the first harmonic of the ECDI in the FFT spectrum (Figure 7).

In the $0.75B_0$, $1B_0$ and $1.25B_0$ cases, the DMD discovered both the ECDI- and MTSI-like modes. In these cases, as well as in the $1.5B_0$ condition, there are other modes of different spatial structures, which feature fluctuations with both radial and azimuthal wavenumber components. As pointed out in part 1 [13], these modes may represent the intermediate stages during which the dominant instability modes are transitioning to one another. Moreover, the long-$\lambda$ mode is identified in the cases with $1.25B_0$ and $1.5B_0$ with the frequency of about 1.5 MHz.

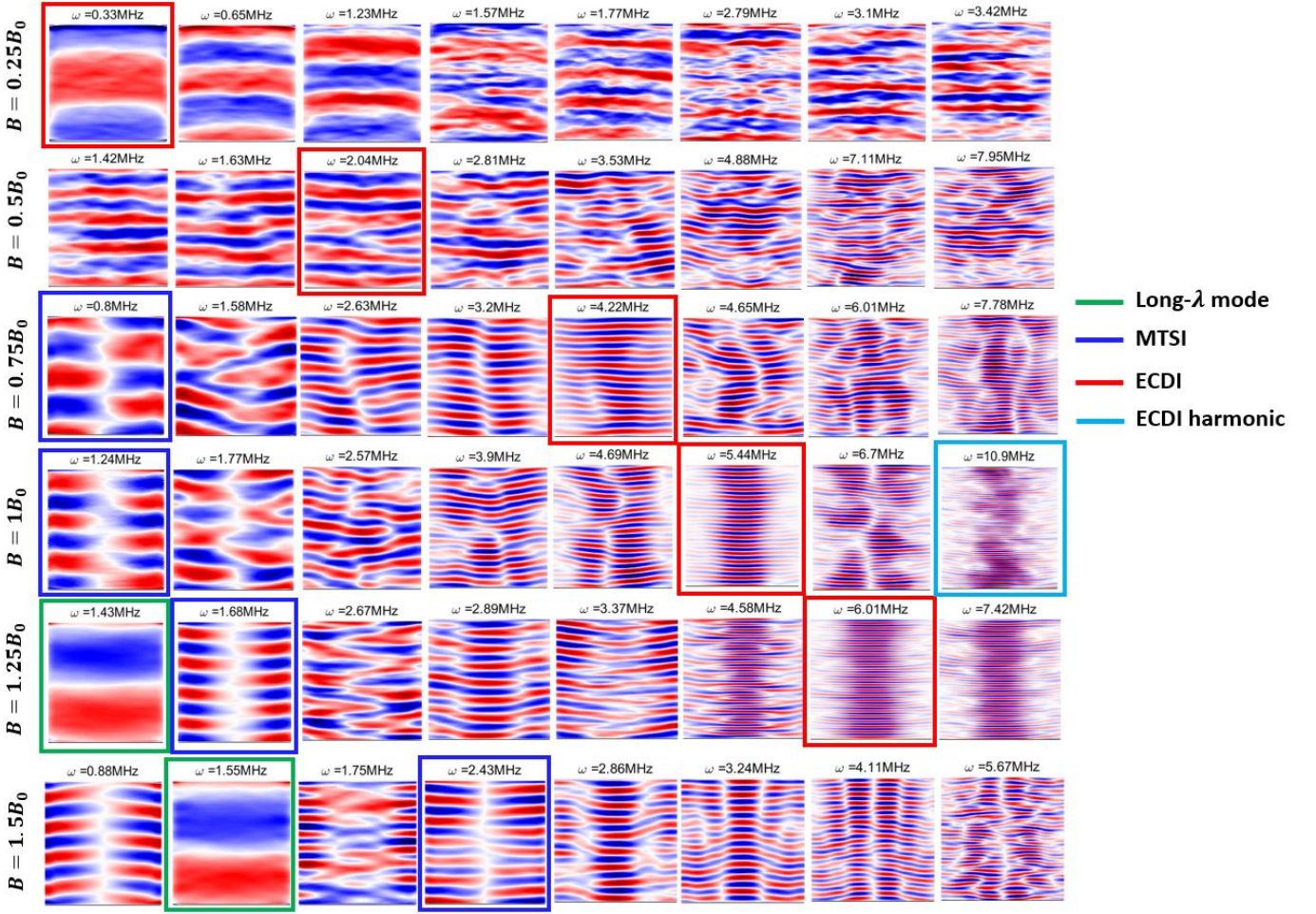

Figure 9: Visualization of the several dominant DMD modes of the azimuthal electric field data from the radial-azimuthal simulations with different $B_x$ values and the xenon propellant. The approach pursued to derive these modes are explained in detail in Ref. [32]. The modes identified with a box correspond to those with wavelengths equal to the wavelengths of the dominant modes visible in the spatial FFT plot in Figure 7.

The comparison of the spatial modes of the ECDI and the MTSI for different propellants as obtained from the OPT-DMD analysis are provided in Figure 10. In each case, the modes' shape is nearly identical among different propellants; however, they are different in terms of the associated frequencies.

To sum up the characteristics of the ECDI and the MTSI as obtained from the FFT and the DMD analyses, we have plotted in Figure 11 for the three propellants the variation in the simulated azimuthal wavenumber, frequency, and the azimuthal phase velocity of the two instabilities vs $B_x$. The analytical relations for the wavenumber of the ECDI's first harmonic [34] and the MTSI's fastest growing mode [35], given by Eqs. 1 and 2, respectively, are superimposed on the wavenumber vs $B_x$ plots for comparison. In Eqs. 1 and 2, $e$ is the elementary charge, $m_e$ is the electrons' mass, and $k_x$ is the radial wavenumber. More detailed discussions about these equations can be found in part I of the article [13].

$$k_z = n\frac{\Omega_{ce}}{V_{d_e}} = n\frac{e}{m_e}\frac{B_x^2}{E_y} \; ; \; n = 1, 2, ..., \tag{Eq. 1}$$

$$k_z = \sqrt{\frac{e}{m_e}\frac{B_x^2}{E_y}k_x}. \tag{Eq. 2}$$

The relations in Eqs. 1 and 2 show that the azimuthal wavenumber of either the ECDI or the MTSI does not depend on the ion mass. Hence, the plotted wavenumbers in Figure 11 (top row) are shared among all propellants.

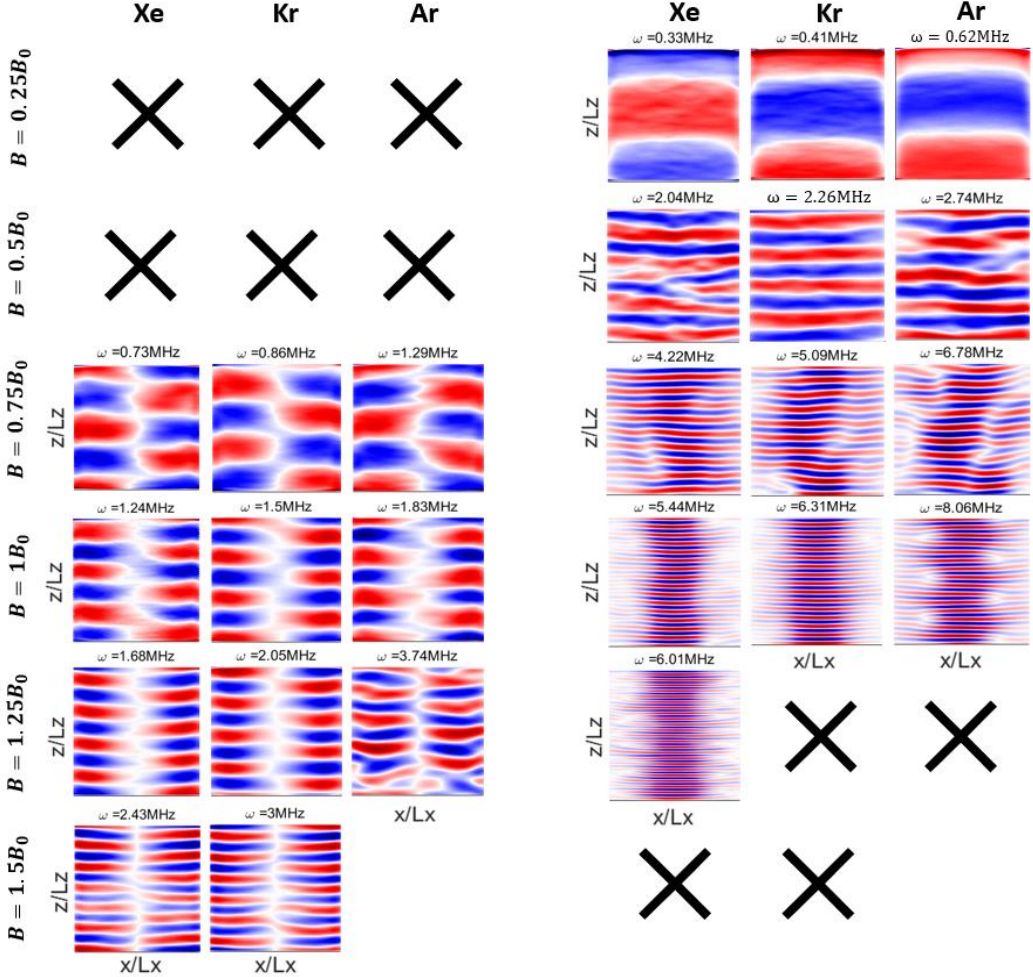

Figure 10: Visualization of the dominant MTSI (left) and ECDI (right) modes from the DMD of the azimuthal electric field data from the radial-azimuthal simulations with different $B_x$ values and the three propellants.

From the top-row plots, we observe that the wavenumbers of both the ECDI and the MTSI increase with the increasing magnetic field. The $B$-field dependency of the ECDI's wavenumber from the simulations is in perfect agreement with the theory, which indicates a quadratic relationship with $B_x$. However, the azimuthal wavenumber of the MTSI coincides with the theoretical relation only at $B_x = 15$ mT and to some extent at $B_x = 20$ mT. Whereas, the simulated MTSI wavenumbers progressively deviate further from the theory as the magnetic field strength increases beyond the range of 15-20 mT. Therefore, although we showed in part I [13] that the analytical relation introduced by Ref. [35] effectively describes the relationship between the MTSI's wavenumber and the axial electric field ($E_y$), it appears inadequate in establishing a precise correlation with $B_x$. The observed discrepancy can be due to the simplifying assumptions made to derive the simple analytical expression in Eq. 2. Nevertheless, further analysis into the exact root cause of this discrepancy is beyond the scope of this effort and is, thus, left for future work.

The frequency of the ECDI and the MTSI (Figure 11 (middle row)) show an increasing trend with the $B$-field. The variation of the instabilities' frequency vs $B$-field magnitude is more pronounced for Ar compared to Kr and Xe. Moreover, for all propellants, the dependency of the ECDI's frequency on the $B$-field appears to flatten out toward larger magnetic fields, whereas the MTSI's frequency is seen to maintain an approximately linear relationship with the $B$-field for the Xe and Kr propellants across the magnetic field range over which it excites. However, the variation of the frequency of the MTSI vs $B_x$ for Ar deviates from a linear behavior at 25 mT.

Finally, looking at the bottom row of Figure 11, the phase velocities of the ECDI and the MTSI are found to be comparable. The ECDI's phase velocity for all propellants reaches its maximum at 10 mT and drops afterwards. The MTSI's phase velocity indicates quite a more moderate variation across the simulated magnetic field intensity range, except in the case of Ar for which the phase velocity sharply rises at $B_x$ = 25 mT.

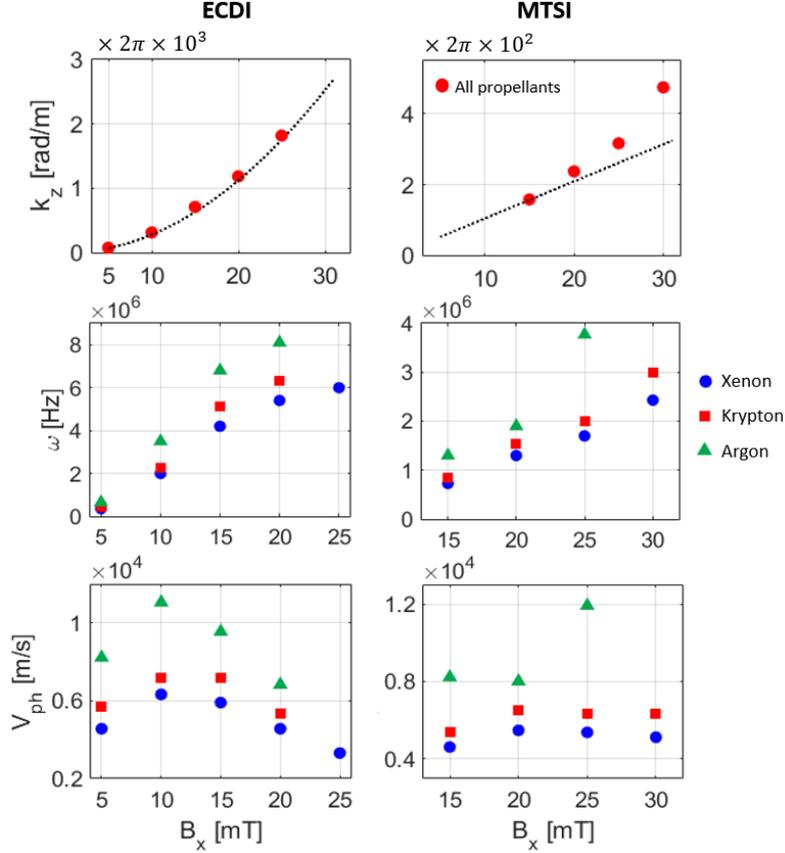

Figure 11: Variation vs $B_x$ of the characteristics of the dominant ECDI and MTSI modes from the radial-azimuthal simulations with different propellants. (Top row) variation of the azimuthal wavenumber ($k_z$) compared against the theoretical relations (Eqs. 1 and 2) for the wavenumber of the ECDI's first harmonic and the MTSI's fastest growing mode shown by black dotted lines, (middle row) variation of the real frequency ($\omega$), (bottom row) variation of the azimuthal phase velocity ($V_{ph}$).

### 3.3. Variation in the electrons' cross-field transport and the species' distribution function

In this section, we compare the significance of the roles played by various electron current carrying mechanisms. These mechanisms are represented by the individual force terms in the electrons' azimuthal momentum equation. This equation was showed in part I of this article [13] to include the convective inertia force ($F_I$), a force due to the non-diagonal pressure terms or viscosity ($F_\Pi$), and the electric force term ($F_E$), which accounts for the momentum change induced by the azimuthal instabilities. Additionally, the magnetic force term ($F_B$), which appears on the left-hand side of the azimuthal momentum equation represents the aggregate effect of all right-hand-side force terms mentioned above. The time-averaged (over 10 $\mu s$) radial profiles of all these force terms and their average values across the radial extent of the domain are presented in Figure 12 and Figure 13, respectively, for various $B$-field simulations with Xe propellant.

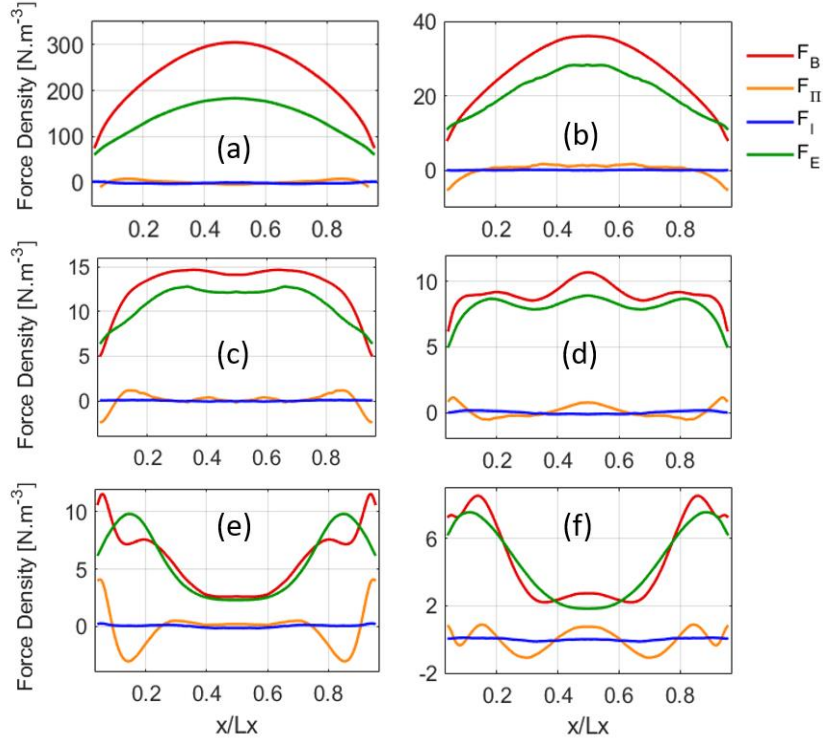

Figure 12: Radial distribution of the force terms in electrons' azimuthal momentum equation (see part I [13]) for the xenon propellant and various values of the radial magnetic field: (a) $0.25B_0$, (b) $0.5B_0$, (c) $0.75B_0$, (d) $1B_0$, (e) $1.25E_0$, (f) $1.5B_0$. The momentum terms are averaged over $10\ \mu s$ of the simulations' time.

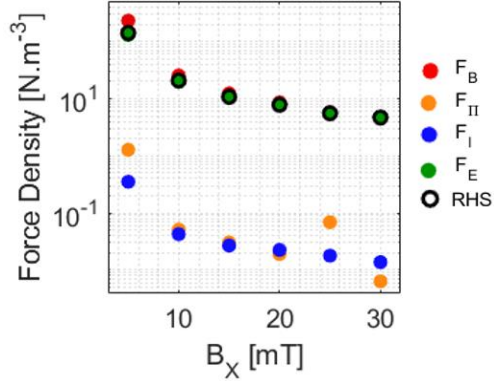

Figure 13: Variation vs $B_x$ of the radially averaged value of each force term in the electrons' azimuthal momentum equation (see part I [13]) from the radial-azimuthal simulations with xenon propellant. The hollow black dot represents the sum of all terms on the right-hand-side of the azimuthal momentum equation.

From these two figures, it is evident that the $F_E$ term is distinctly dominant in all cases such that its average value across different $B_x$ intensities is more than 2 orders of magnitude larger than the value of the other terms. The distributions of the $F_B$ and the $F_E$ terms in the $0.25B_0$ and $0.5B_0$ cases show a convex profile with the maximum occurring at the center. This is consistent with the transport profiles we expect when the ECDI is the dominant instability in the radial-azimuthal configuration. In the $0.75B_0$ and $1B_0$ cases, the superposition of the contributions of the ECDI and the MTSI to electron transport creates a more radially uniform $F_E$ distribution. Particularly in the $1B_0$ case where the MTSI and the ECDI are of more comparable intensities, the $F_E$ profile contains two peaks away from the center indicating the MTSI's transport and a peak at the center which represents the transport carried mainly by the ECDI. Finally, the profiles of the $F_E$ term for the $1.25B_0$ and $1.5B_0$ conditions show peaks near the edges of plasma as a result of the strong MTSI contribution.

The role of the convective inertia ($F_I$) term is negligible in all cases. The viscous term ($F_\Pi$) is most significant in the $1.25B_0$ and $1.5B_0$ conditions. In this respect, although the radial average of $F_\Pi$ term is still small compared to $F_E$ in these cases, the local value of this term along the radius has a noticeable effect on the total momentum distribution captured by $F_B$ in the high magnetic field intensity conditions. In fact, in the $1.25B_0$ case, $F_\Pi$ accounts for nearly two-thirds of the total momentum ($F_B$) adjacent to the walls and it becomes negative immediately after the near-wall region toward the center of the domain where it counteracts the $F_E$ term and, thus, reduces the $F_B$. The $F_\Pi$ term then assumes a nearly constant and slightly positive value in the central part of the domain. In the $1.5B_0$ condition, the $F_\Pi$ demonstrates an oscillatory behavior across the domain reaching its peak relative significance at the centerline where it creates a bump in the $F_B$ profile.

As one last point concerning the plot in Figure 13, the discrepancy between the left-hand side ($F_B$) and the sum of the force terms on the right-hand side (RHS) of the azimuthal momentum equation from the radial-azimuthal simulations at low $B_x$ intensities is due to the momentum loss induced in the system by the introduction of an artificial boundary along the axial direction. More details on this is provided in part I [13].

Finally, to investigate how the instabilities affect particles' velocities, we have provided the radial and azimuthal velocity distribution functions of the electrons and the ions at t = 30 $\mu s$ in Figure 14 and Figure 15, respectively.

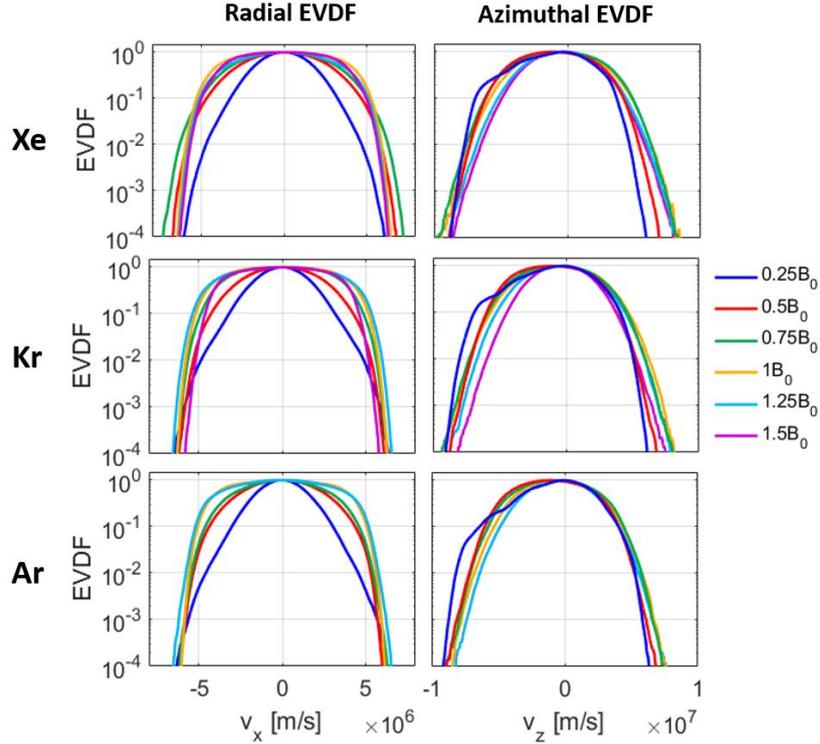

Figure 14: Normalized radial and azimuthal velocity distribution functions of the electrons for various radial magnetic field values and the three studied propellants from the quasi-2D simulations.

Looking at the radial electron VDFs, we notice that, at the lowest magnetic field (the $0.25B_0$ case), the distribution of particles is closer to a Maxwellian for all propellants. However, in the other cases, the high-velocity tail of the distribution functions is depleted, and most of the particles have low-to-mid range velocities. Furthermore, the azimuthal EVDFs become increasingly asymmetric with respect to the $v_z = 0$ axis as the $B$-field decreases. In particular, the azimuthal EVDF in the $0.25B_0$ case show a noticeable peak at the negative velocities, which represent a large population of drifting electrons at relatively high azimuthal drift velocities compared to the other cases.

Referring to the ions' velocity distribution functions in Figure 15, we observe that, in low $B$-field range ($0.25B_0$ and $0.5B_0$), the radial IVDFs have lost the mid-range velocity ions and retained the ions with high velocity in the tail. In the higher $B$-field intensity conditions, the mid-range velocity ions are seen to be replenished.

The azimuthal IVDFs clearly show the ions' interaction with the azimuthal waves. Indeed, the case with $B_x = 0.25B_0$ demonstrates a strong broadening, which becomes increasingly more significant from Xe to Ar. Unlike the other $B_x$ conditions, and even unlike what was observed for various $E$-field cases presented in part I [13], the azimuthal IVDF in the $0.25B_0$ case broadens toward both the negative and the positive velocities. This could be possibly because some of the azimuthal ECDI waves in this case may have positive phase velocities which can, in turn, cause the interacting ions to gain azimuthal velocities in the positive direction as well. Indeed, we had pointed out from the DMD plots in Figure 9 that the azimuthal modes in the $B_x = 0.25B_0$ case seem rather turbulent.

The extent of IVDFs' broadening in the other $B_x$ cases are almost similar with minor differences across the different propellants. This is consistent with the small variation in the phase velocities of the azimuthal waves across the simulated $B$-field range as was observed in the bottom row plots of Figure 11.

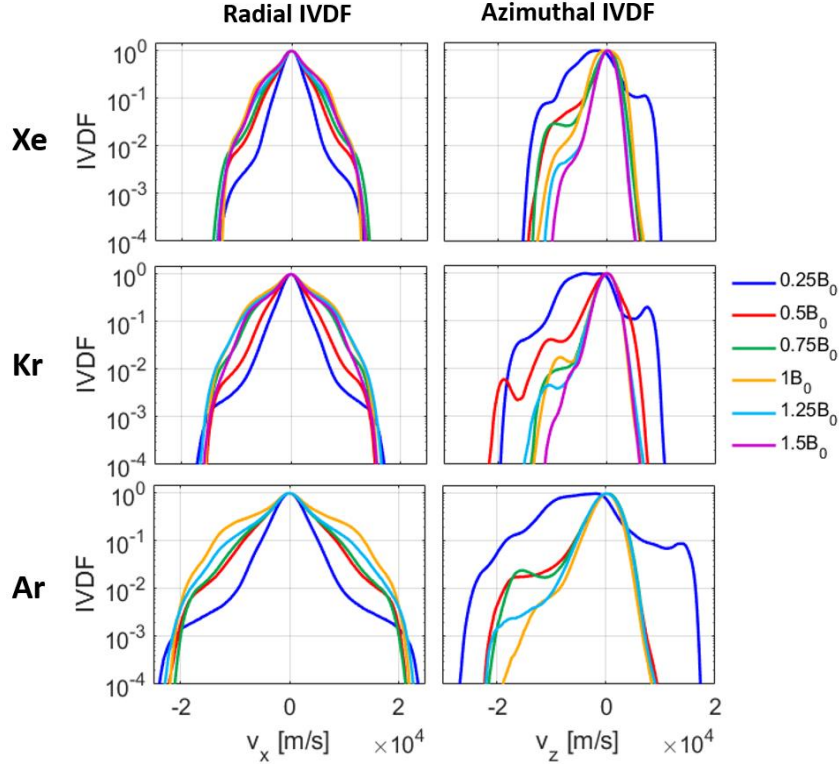

Figure 15: Normalized radial and azimuthal velocity distribution functions of the ions for various radial magnetic field values and the three studied propellants from the quasi-2D simulations.

### Section 4: Conclusions

The studies in this paper complemented the discussions of part I [13] by assessing in depth the effects that the radial magnetic field intensity has on the behavior of the plasma discharge of various gases, namely, xenon, krypton, and argon. Similar to part I [13], we carried out the investigations in a 2D configuration representative of a Hall thruster's radial-azimuthal cross-section and using the reduced-order quasi-2D PIC with a uniform 50-region domain decomposition approximation. The magnetic field value was varied over the range of 5 to 30 mT, with 5 mT increments. Having considered the $B$-field intensity of 20 mT as the baseline value ($B_0$), the studied $B$-field range corresponded to $0.25B_0$ to $1.5B_0$. We demonstrated that the variation in the magnitude of the magnetic field yields three plasma regimes across all propellants.

In regime I associated with relatively low $B$-field values (0.25 to $0.5B_0$), we observed the absence of the MTSI mode, and, as a result, a radially averaged electrons' radial-to-azimuthal temperature ratio ($T_{ex}/T_{ez}$) of less than one and a radial distribution of the axial electron current density that peaked around the center of the domain due to the dominance of the ECDI mode. In regime II with moderate values of the magnetic field intensity (0.75 to $1B_0$), the MTSI develops and presents a comparable strength compared to the ECDI. The mean $T_{ex}/T_{ez}$ ratio in this regime was accordingly observed to be close to 1, and the axial electron current density exhibited a rather uniform radial profile. An exception was observed for Ar propellant for which, at the $B$-field intensity of $1B_0$, the MTSI was observed to stronger than the ECDI, which resulted in the mean $T_{ex}/T_{ez}$ ratio of slightly larger than 1 and a radial $J_{ey}$ profile that peaked near the walls. For relatively high values of the $B$-field (1.25 to $1.5B_0$) corresponding to regime III, the ECDI was overall observed to be absent, with the instabilities' spectra dominated by the MTSI and a long-wavelength wave mode. For regime III, $\frac{T_{ex}}{T_{ez}} > 1$, and the radial profile of the $J_{ey}$ has peaks in the vicinity of the walls. In case of Xe propellant, it is noteworthy that the ECDI was observed to still exist at $B_x = 1.25B_0$ although at a mitigated strength.

Another particularly interesting influence of the magnetic field intensity on the plasma properties was noticed from the radial profiles of ion Mach number. Indeed, we observed that, as the $B$-field intensity increases, the ion sonic point shifts toward the plasma bulk away from the walls. It was, thus, concluded that, at $B$-field values corresponding to the plasma regimes II and III, the Bohm criterion is no longer an adequate measure to delimit the edge of the sheath.

The detailed FFT and DMD analyses of the azimuthal electric field fluctuations confirmed the observations summarized above and inferred from the mean $T_{ex}/T_{ez}$ ratio as well as the radial profile of the $J_{ey}$ concerning the nature of the dominant instability modes within the three plasma regimes. The DMD analyses in particular casted light on the spatial structure of the dominant modes that were identified through the FFT plots and were seen to notably aid in disambiguating the results of the spatial and temporal FFT analyses.

In regard to the aggregated insights from the characterization of the instabilities, we first showed that the dependency of the azimuthal wavelength of the ECDI's first harmonic on the $B$-field intensity from the simulations matches well the theoretical relationship. However, we noticed that the simplified relation proposed by Ref. [35] for the azimuthal wavelength of the MTSI's fastest growing mode does not capture the correct correlation with the magnetic field intensity, especially for the value ranges corresponding to regimes II and III. Second, the frequencies of the ECDI and the MTSI were noticed to increase with $B_x$, with the increase being more pronounced for Ar compared to Xe and Kr. Third, regarding the phase velocity vs $B_x$ for the ECDI and the MTSI, we noticed a non-monotonic increasing-then-decreasing trend in case of the ECDI for all propellants. However, in case of the MTSI, even though the trend for all propellants was observed to be rather contact across the magnetic field intensity range in regime II, Ar presented a sharp rise in phase velocity in regime III at the $B_x$ value of $1.25B_0$.

Concerning the variation in the significance of the force terms in the electrons' azimuthal momentum equation, we discussed, as an example for xenon, that the electric force term representing the effect of the instabilities is the dominant contributor to transport across the magnetic field values investigated. The radial distributions of the $F_E$ term and the $F_B$ term (the left-hand side of the momentum equation) were noticed to expectedly change with $B_x$ value in accordance with the observations reported from the time-averaged profiles of the axial electron current density. The viscous force term ($F_p$) was seen to have a non-negligible contribution to total electron cross-field transport as represented by the $F_B$ term within regimes II and III along the increase in the $B_x$ intensity.

Finally, we evaluated the impact of the change in plasma regime due to the $B_x$ intensity increase on the velocity distribution functions of the electrons and the ions. Regarding the electrons, we noted that, with the increase in $B_x$ and for all propellants, the radial EVDFs depart from a nearly Maxwellian distribution at the lowest $B$-field intensity to distributions where the tail of the VDF is consistently depleted. Whereas, the azimuthal EVDF became increasingly more symmetric with respect to $v_z = 0$ axis with the rise in $B_x$ value.

Regarding the ions, the extent of distortion and broadening of the radial and azimuthal VDFs, respectively, were seen to decrease with the increase in the magnetic field magnitude. An observation of particular interest was that the ions' azimuthal VDF featured a broadening in both the positive and negative velocity directions at $B_x = 0.25B_0$. This was the case for all propellants albeit the extent of broadening was different among various gases. The two-sided IVDF broadening was conjectured to possibly root in the turbulent ECDI spectra existing at this low $B$-field magnitude.

**Acknowledgments**:

The present research is carried out within the framework of the project "Advanced Space Propulsion for Innovative Realization of space Exploration (ASPIRE)". ASPIRE has received funding from the European Union's Horizon 2020 Research and Innovation Programme under the Grant Agreement No. 101004366. The views expressed herein can in no way be taken as to reflect an official opinion of the Commission of the European Union.

The authors gratefully acknowledge the computational resources and support provided by the Imperial College Research Computing Service (http://doi.org/10.14469/hpc/2232).

**Data Availability Statement**:

The simulation data that support the findings of this study are available from the corresponding author upon reasonable request.